\documentclass[twocolumn,showpacs,preprintnumbers,amsmath,amssymb,prb,aps,superscriptaddress,floatfix]{revtex4}
\usepackage{subfigure}
\usepackage{epsfig}
\usepackage{color}
\usepackage{graphicx}
\usepackage{amsmath}

\usepackage{amssymb}
\usepackage{gensymb,bm}%
\usepackage{float}
\usepackage{units}

\begin{document}

\title{Improper origin of polar displacements at CaTiO$_3$ and CaMnO$_3$ Twin Walls}

\author{Paolo Barone}
\affiliation{Consiglio Nazionale delle Ricerche - CNR-SPIN, I-67100 L'Aquila, Italy}

\author{Domenico Di Sante}
\affiliation{Consiglio Nazionale delle Ricerche - CNR-SPIN, I-67100 L'Aquila, Italy}
\affiliation{Department of Physical and Chemical Sciences,
University of L'Aquila, Via Vetoio 10, I-67100 L'Aquila, Italy.}

\author{Silvia Picozzi}
\affiliation{Consiglio Nazionale delle Ricerche - CNR-SPIN, I-67100 L'Aquila, Italy}

\begin{abstract}

Recent interests towards novel functionalities arising at domain walls of ferroic materials naturally call for a
microscopic understanding. To this end, first-principles calculations have been performed in order to provide solid evidence of
polar distortions in the twin walls of nonpolar CaTiO$_3$ and magnetic CaMnO$_3$. 
We show that such polar displacements arise from rotation/tilting octahedral distortions ---   cooperatively acting at the twin wall
in both considered systems --- rather than from a proper secondary ferroelectric instability, as often believed.
Our results are in excellent agreement with experimental observations of domain walls in CaTiO$_3$. In addition, we show
that magnetic properties at the twin wall in CaMnO$_3$ are also modified, thus suggesting an unexplored route to achieve and detect
multiferroic ordering in a single-phase material.

\end{abstract}


\maketitle

\section{Introduction}
For long time, domain boundaries in ferroelectric and ferroelastic materials
have been looked at as mere juxtapositions of materials in the bulk state, 
lacking any interesting physical
meaning on their own. 
In the last decade their role has been reconsidered, leading to the realization
that domain walls (DW) can display
novel features which do not emerge in the bulk. 
As such, they can become active elements of
potential new devices, leading to the``domain wall engineering''
concept\cite{Wada2006,Salje2010,Salje_annrev,Catalan2012}.
The emergence of new functionalities at domain walls (DW) can be loosely understood in the
framework of Landau theory of phase transitions. In fact, walls separate domains that are characterized by a
primary order parameter (OP) pointing in two or more directions (polarities), implying that some component of the OP
must monotonically
decrease when approaching the domain boundary and eventually vanish at the wall.
As a consequence, competing secondary order parameters,
suppressed in the bulk, may in principle emerge in proximity of the boundaries
\cite{Lee_Salje_Bismayer1,Lee_Salje_Bismayer2,Salje}.

A paradigmatic example of the competition between primary and
secondary OPs has been put forward by the ferroelastic perovskite
CaTiO$_3$\cite{Goncalves-Ferreira_Redfern_Artacho_Salje}. This material  crystallizes at ambient
conditions in a nonpolar $Pnma$ structure ($Pbnm$ setting  adopted in the following)
characterized by
antiferrodistortive distortions (AFD), which can be described as oxygen-octahedra tiltings and rotations,
accompanied by a spontaneous ferroelastic strain. 
The parent cubic
structure exhibits a secondary ferroelectric (FE) instability, namely a polar off-centering of Ti within the oxygen
octahedral cages associated to the FE polarization\cite{Cockayne_Burton}.
Particularly interesting is the prediction of FE instabilities in
magnetic $A$MnO$_3$ ($A$ = Ca, Sr, Ba)\cite{Ghosez_Bousquet, Barone_CMO, rondinelli_prb, Bersuker}
which questioned the usually invoked empirical ``exclusion rule'' for magnetic and FE
perovskites\cite{spaldin_hill}, as   experimentally confirmed in (Sr,Ba)MnO$_3$\cite{sakai} and
strained CaMnO$_3$ thin films\cite{cmo_film}. Therefore, the possibility of  FE twin walls in this class
of magnetic compounds may represent a new and unexplored route to achieve nanoscale multiferroic features in single-phase
materials. Furthermore, a significant magnetoelectric coupling may be expected, since both magnetism and FE
features would originate from the very same B-site ions.

In principle, the presence of twin ferroelastic domains in CaTiO$_3$ could imply a suppression of the primary AFD OP
at the domain wall and a consequent activation of the secondary FE instability. This motivated recent 
numerical calculations based on an atomic-scale, though empirical, description
of the wall, predicting ferrielectricity at CaTiO$_3$ twin boundaries,
with maximum dipole moments appearing at the wall.\cite{Goncalves-Ferreira_Redfern_Artacho_Salje}
Very recently, a direct observation of such ferrielectric domain boundaries in CaTiO$_3$
was achieved by transmission electron microscopy
\cite{VanAert_Turner_Delville_Schryvers_VanTendeloo_Salje},
reporting, however, offcentric displacements
one order of magnitude larger than those theoretically predicted.
More importantly, the simple interpretation based on the coexistence/competition of primary and secondary OPs may not really apply to CaTiO$_3$, since
the AFD distortions never fully disappear in the experimentally observed twin walls,
while the polar offcenterings have
been found to be strongly locked to the twin angle and local pattern of AFD distortions. 
These observations apparently
suggest an improper origin for the emerging DW ferroelectricity, hindering the possibility of switching its polarization via an applied electric field. On the other hand, a sizeable DW electric polarization could serve in principle as an additional handle for, e.g., controlling twin-walls dynamics.


Motivated by these premises, we performed an accurate analysis of DW in
both CaTiO$_3$ (CTO) and CaMnO$_3$ (CMO) in the framework of first-principles density-functional-theory (DFT) calculations.
We find that polar offcenterings are caused by the peculiar interplay of cooperative AFD rotations at the wall, rather than by the activation of a secondary FE instability. Microscopically, the mechanism can be viewed as a local realization of the recently proposed hybrid improper ferroelectricity
\cite{benedek_prl,Jorge,rondinelli_advmat}. In this framework, the AFD pattern of tilting/rotations of $B$O$_6$ octahedra is expected to cause antipolar offcenterings of $A-$site cations\cite{Jorge,rondinelli_advmat}; as a consequence, twin boundaries in perovskite oxides could play the role of walls between domains with different antipolar distortions, implying in principle a non-negligible contribution of $A-$site cations to the wall polarization. On the other hand, a significant contribution to polarization is predicted to originate as well from B-site ions, that offcenter significantly even in the presence of a magnetic ordering (an unexpected effect for CMO at ambient conditions) in response to the local pattern of AFD distortions at the DW.

\section{Methods and computational details}

We adopted the PBEsol generalized-gradient approximation for the exchange-correlation
functional revised for solids
\cite{Perdew_Ruzsinszky_Csonka_Vydrov_Scuseria_Constantin_Zhou_Burke} as implemented in the VASP code\cite{vasp},
using a 500 eV plane-wave cutoff and
a 1$\times$4$\times$4 Monkhorst-Pack mesh.
Cell and ionic relaxations have been performed until forces acting on ions were smaller than 0.01 eV/\AA. 
Furthermore, as CMO is a G-type antiferromagnetic (AFM-G) insulator in its orthorhombic
ground-state\cite{Matar}, we impose the AFM-G
spin-ordering away from the  wall, while allowing for different types of parallel/antiparallel-spin bonds at the
interface.
\begin{figure}[b]
\begin{center}
\includegraphics[width=0.48\textwidth,angle=0,clip=true]{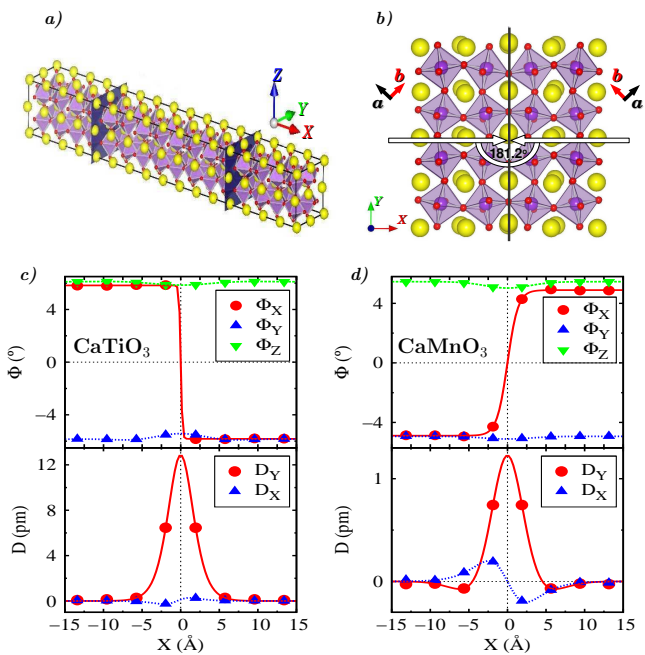}
\caption{(a) Ball-and-stick model of CaTi(Mn)O$_3$ supercell. (b) Top view highlighting the ferroelastic domains, with
orthorhombic cells mirrored with respect to the $[1\bar{1}0]$ plane, and
the twin angle of CaTiO$_3$. Yellow and red balls refer to Ca and O atoms, respectively, while
Ti (Mn) ions are inside purple octahedra. (c)-(d) Layer-averaged rotational order parameters as a function of the distance from twin
wall and associated offcentering of B-site ions estimated from the center-of-mass of oxygen octahedra for (c) CTO
and (d) CMO. The DW profile of the polar offcenterings was fitted via a function
$D_y(X) = D_0\, \text{sech}^2\left(\nicefrac{X}{\xi_D}\right)$\cite{Salje}.}
\label{fig1}
\end{center}
\end{figure}
Since two DW are needed to implement periodic boundary conditions,
large supercells have been built comprising 16 pseudocubic $AB$O$_3$ unit cells along a direction $X$ perpendicular to the wall
and 2 unit cells along directions $Y$ and $Z$ parallel to the wall, for a total
of 320 atoms.
Different domains may be
identified by defining the primary rotational order parameter as an axial vector $\Phi$
from the static rotational
momenta $\bm \phi (\bm R) \propto \sum_{l=1,6} \hat{\bm r}_l \times \hat{\bm r}_l^{'}$\cite{Gopalan}. Here,
$\bm R = i\bm a_X+j\bm a_Y+k\bm a_Z$ is the $B$-site supercell lattice vector, while
$\hat{\bm r}_l$ ($\hat{\bm r}_l^{'}$) represents oxygen positions within
each $B$O$_6$ octahedron before (after) the static rotation about an axis passing through
the $B$O$_6$ centre of mass. The cartesian components $x,y,z$ of the axial vector $\bm \phi (\bm R)$
correspond to rotations about corresponding axes $X,Y,Z$. The
order parameter for each layer parallel to the DW is then defined by
including  appropriate phase factors  as:
\begin{eqnarray}
\Phi_{x,y}(X)&=&(-1)^i\,\frac{1}{4}\,\sum_{j,k}(-1)^{j+k}\phi_{x,y}(\bm R) \\
\Phi_{z}(X)&=&(-1)^i\,\frac{1}{4}\,\sum_{j,k}(-1)^{j}\phi_{z}(\bm R),
\end{eqnarray}
corresponding to the general $a^-b^-c^+$ rotational pattern in
Glazer's notation\cite{Glazer}.


\section{Results}
\subsection{Domain wall structure and polarization profile}
After ionic relaxations, the energetically more favorable DW is a ferroelastic
wall obtained through a mirror twin law about the $(1\bar{1}0)$ plane of the $Pbnm$ structure, in agreement with the experimental observations\cite{VanAert_Turner_Delville_Schryvers_VanTendeloo_Salje},
with an estimated DW energy $E_{DW}= 16\, (41)\, mJ/m^2$ for CTO (AFM-G CMO)\cite{note2}.
The twin wall is characterized by a switching of
the out-of-phase BO$_6$ rotations around the supercell $X$ axis at the wall\cite{footnote_extra} as described by
$\Phi_x(X)\propto\tanh(\nicefrac{X}{\xi})$ [see Fig. \ref{fig1}]. 
After cell relaxation, a twin angle of 181.2$^\circ$ is found for CTO, in excellent agreement with
experimental findings\cite{VanAert_Turner_Delville_Schryvers_VanTendeloo_Salje}. In Fig. \ref{fig1}(c)-(d) we show
the layer-averaged offcentering $\bm D^{\mbox{\footnotesize Ti(Mn)}}=\nicefrac{1}{4}\,\sum_{j,k}\,{\bm d}^{\mbox{\footnotesize Ti(Mn)}}(\bm R)$ of Ti (Mn),  where ${\bm
d}^{\mbox{\footnotesize B}}(\bm R)=\nicefrac{1}{6}\sum_{l=1,6}\left({\bm r}^{B}
-{\bm r}^{O}_l\right)$ describes the local displacement of B-site ions with respect to the center of mass of each
$B$O$_6$ octahedron. Two ion offcenterings
clearly appear at the wall: a polar distortion along DW direction $Y$ as large as $\unit[6.5]{pm}$ ($\unit[0.7]{pm}$) 
for CTO (AFM-G CMO), and an antiphase (odd) polarization developing perpendicularly to the wall.
Interestingly, if we define the Ti (Mn)
offcenterings with respect to Ca sites (as done in Ref. \cite{VanAert_Turner_Delville_Schryvers_VanTendeloo_Salje}), the
sign of the Ti (Mn) displacement along $Y$ is reversed, amounting to $\simeq-\unit[6.2]{pm}\,(-\unit[3.1]{pm})$. If on one side the magnitude of the effect is again in
excellent agreement with the experimental value $\unit[6.1]{pm}$, the change of sign suggests that actually also Ca ions are displaced with
respect to oxygens. Indeed, one can also define the Ca offcentering from the center of mass of a dodecahedral cage
$A$O$_{12}$ as $\bm d^{\mbox{\footnotesize Ca}}(\bm R)=
\nicefrac{1}{12}\sum_{l=1,12}\left({\bm r}^{\mbox{\footnotesize Ca}}
-{\bm r}^{\mbox{\footnotesize O}}_l\right)$, finding a layer-averaged offcentering of Ca with respect to the Os as large as $\unit[21.7]{pm}$ ($\unit[6.8]{pm}$)  along the $Y$ direction at the CTO (AFM-G CMO) DW.

\begin{figure}[b]
\begin{center}
\includegraphics[width=0.48\textwidth,angle=0,clip=true]{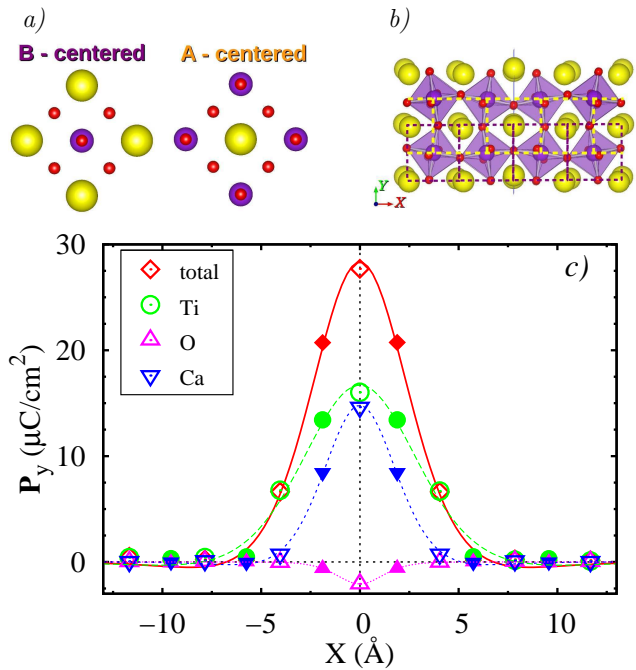}
\caption{a) Sketch of the two different choices of unit cells used for the calculation of polarization
via Born effective charges. The polarization profile in the supercell is then evaluated as sketched in b).
c) Total polarization profile at CTO twin wall and contributions assigned to different ions; empty and solid symbols refer to different choices of the unit cell, referring respectively to $A$-site and $B$-site centered cells.
}
\label{fig2}
\end{center}
\end{figure}
In order to get a deeper insight on the distortions close to the twin boundary, we evaluated the polarization profile from the unit-cell polarization 
\begin{equation}\label{eq:pol}
\bm P^{(i)} =\frac{e}{\Omega_c}\sum_\alpha\,w_a \bm Z^*_\alpha\cdot \bm u_\alpha^{(i)}.
\end{equation} 
Here $e$ is the electron charge, $\Omega_c$ is the volume of a five-atom (bulk) unit cell, $\bm Z^*_\alpha$ are the Born effective charge tensors and $\bm u_\alpha^{(i)}$ describe the ionic displacement of ion $\alpha$ from its bulk position
in unit cell $i$.\cite{vanderbilt_PTO}
The coefficients $w_\alpha$ account for the possible overcounting of ionic contributions, since some of the ions in the five-atom unit cell are shared by neighboring cells, and depend on the choice of the unit cell. We adopted two choices [shown in Fig. \ref{fig2}(a)], namely a $B$-site centered, with a Ti (Mn) ion sitting in the center of the unit cell, and a $A$-site centered cell, where a Ca ion is located at the center of the cell. In the first case, each of the six neighboring oxygens is shared by two unit cells, whereas each Ca ion is shared by eight unit cells, implying $w_O=1/2$ and $w_{Ca}=1/8$ (weight factors for $A$-centered cells are analogously chosen as $w_O=1/4$ and $w_{B}=1/8$).
Within this formulation, the origin of the reference frame can be arbitrarily chosen, since only relative displacements are taken into account; furthermore, the contribution to total $\bm P$ due to each individual ion can be unambigously identified, corresponding to  Eq. (\ref{eq:pol}) where only
terms relative to the chosen ion are included. 
In Fig. \ref{fig2}(a) we show the layer-averaged polarization profile for a CTO DW, where both Ca and Ti ions are shown to contribute significantly to the total $\bm P$, resulting in a DW polarization as large as $27.7~ \mu C/cm^2$. 

\subsection{Improper origin of domain wall polarization}
\begin{figure}[t]
\begin{center}
\includegraphics[width=0.44\textwidth,angle=0,clip=true]{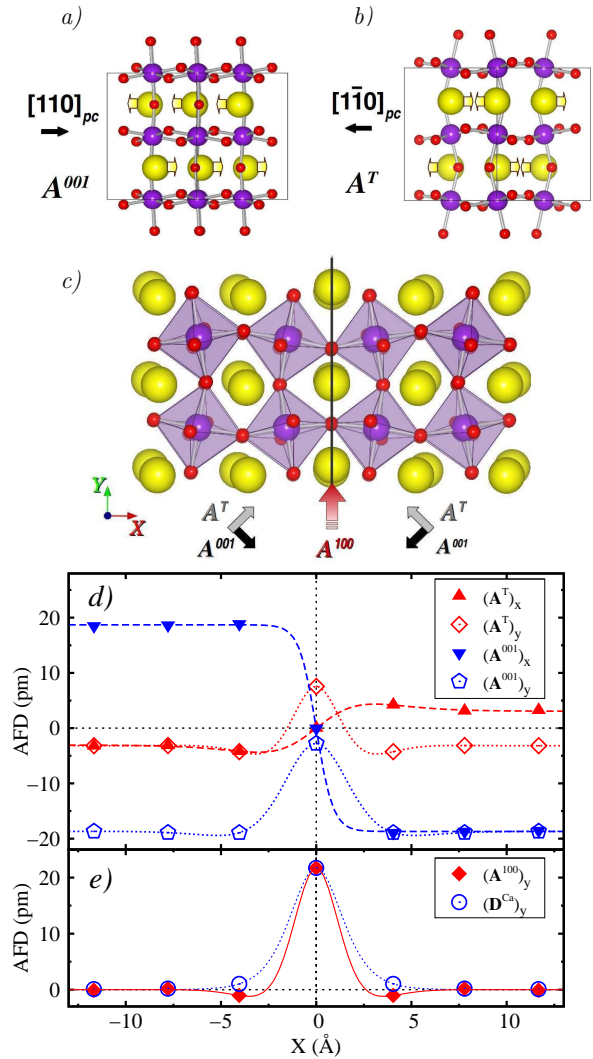}
\caption{Antipolar displacements of Ca ions in the bulk $Pbnm$ structure of Ca$B$O$_3$, showing the two
kind of modulated collective antiferrodistortive distorsions a) ${\bm A}^{001}$ and b) ${\bm A}^T(X)$.
c) Antipolar displacements at the considered twin wall, separating two twin domains characterized
by different AFD polarizations; the emergence of the third local displacement induced by the local rotational pattern is also highlighted.
d) DW profile of bulk antipolar order parameters in CTO, describing antipolar displacements of Ca ions evaluated in
the local centre of mass of each CaO$_{12}$ dodecahedron. e) Layered-averaged polar offcentering $\bm D^{\mbox{\footnotesize Ca}}$ of Ca ions,
compared with the DW profile of the antipolar OP ${\bm A}^{100}$ that is expected to develop only at the wall. 
}
\label{fig3}
\end{center}
\end{figure}
The Ca contribution to $\bm P$ can be understood in terms of coupled AFD rotational modes. By taking into account symmetry-allowed trilinear
couplings with general expression $\Phi_\alpha\Phi_\beta D_\gamma$, in fact, it has been shown that the $a^-a^-c^+$ pattern
adopts two kinds of antipolar displacements of A-site ions\cite{Jorge}. The first type, ${\bm A}^{001}$,
involves A-ion displacements along $[110]_{pc}$ directions of the pseudocubic cell,  modulated in anti-phase
when going from a (001) CaO plane to an adjacent one; in terms of $\bm d^{\mbox{\footnotesize Ca}}(\bm R)$ it can be expressed as:
\begin{equation}\label{eq:a001}
{\bm A}^{001}(X)=\frac{1}{4}\,\sum_{j,k}(-1)^k\,{\bm d}^{\mbox{\footnotesize Ca}}(\bm R).
\end{equation}
The second type,
${\bm A}^T(X)$, occurs along the $[1\bar{1}0]_{pc}$ direction and has an anti-phase
modulation along the three pseudocubic directions, reading:
\begin{equation}\label{eq:at}
{\bm A}^T(X)=(-1)^i\,\frac{1}{4}\,\sum_{j,k}(-1)^{j+k}{\bm d}^{\mbox{\footnotesize Ca}}(\bm R).
\end{equation}
Their profile across the DW is shown
in Fig.\ref{fig2}, suggesting that {the twin boundary
represents in fact a wall between domains with different antipolar polarities}; specifically, the $X$ component of
both the antipolar OPs vanishes at the boundary and is opposite in different ferroelastic domains. 
A third antipolar distortion ${\bm
A}^{100}=(-1)^i\nicefrac{1}{4}\sum_{j,k} {\bm d}^{\mbox{\footnotesize Ca}}(\bm R)$ is also predicted to be induced by the local AFD distortions at the DW\cite{footnote_extra},
involving displacements along the $[010]_{pc}$ direction which are modulated in
antiphase as one moves from a (100) CaO plane to the next one.
As shown in Fig.\ref{fig2}, the major contribution to the Ca displacements comes in fact from this mode.
On the other hand, once the
inversion symmetry is locally broken by the AFD DW, the FE instability of B-site ions is easily activated, {\it but only along the
directions dictated by the primary non-polar distortions (rotations)}  causing the symmetry breaking\cite{note3}. Thus, a
significant displacement of B-site ions occur parallel to the $Y$ axis, while the switching of the $X$ component of the Ca
motions across the wall causes the anti-phase (odd) modulation of $D_x^B(X)$. Such offcentering distortions are tipically smaller than those of Ca ions; it is worth to notice, however,
 that both Ti and Mn show anomalous Born effective charges ($Z^*_{{\text Ti},yy} = +7e$ and $Z^*_{{\text Mn},yy} = +6.8e$), thus contributing significantly to the DW electric polarization. 
 Eventually, the devised improper origin of DW polarization easily explains the observed locking of the ionic offcentering to the twin angle and local AFD rotational pattern. Following Ref. \onlinecite{Jorge}, the most relevant coupling between distortional modes at the wall is found to occur trilinearly between $d_y^{\mbox{\footnotesize Ca}}, \phi_x$ and $\phi_y$. In principle, the polar distortion could be reversed by reversing one of these two rotational momenta, a possibility hardly attained in a realistic case due to the cooperative character of rotational distortions that would require a reversal of the rotational pattern in the whole domain. Nonetheless, this situation is realized in our supercell where the two symmetric DWs are characterized by exactly the same $\phi_y$ (and $\phi_z$) but opposite $\phi_x$, and where the polarization profile is indeed found to be reversed.


\subsection{Magnetic exchange at the domain wall} 
\begin{figure}[!ht]
\begin{center}
\includegraphics[width=0.48\textwidth,angle=0,clip=true]{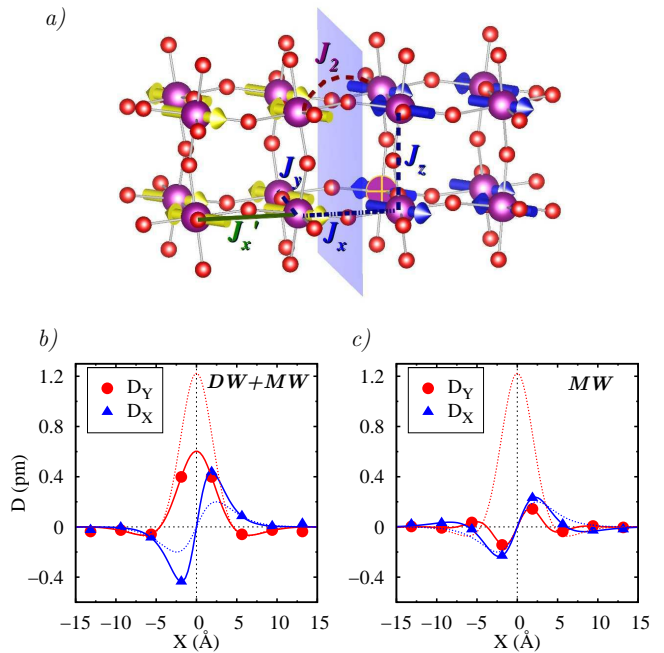}
\caption{(a) Sketch of the evaluated DW exchange couplings at the interface; the domain wall between two AFM-G domains with all parallel-spin bonds across the DW, labeled as FM$_X$ configuration, is also highlighted . (b)-(c) Mn
offcentering profile when the magnetic domain wall
is assumed on top of the ferroelastic twin boundary [(b): DW+MW] and of the bulk lattice structure [(c): MW], compared to
the single AFM-G magnetic domain with ferroelastic twin wall (dotted lines).}
\label{fig4}
\end{center}
\end{figure}

As for CMO, an interesting additional degree of freedom is brought about by the localized magnetic moments on Mn ions. If,
on one hand, all previous considerations perfectly apply to the case of a single magnetic domain (with ground-state AFM-G
configuration), on the other hand significant spin-phonon coupling effects may be expected at CMO DWs.\cite{Hong_Stroppa_Iniguez} We considered then a
selection of possible interface spin configurations, with parallel-spin bonds along $X,Y$ or $Z$ direction. As expected,
all these spin configurations resulted in higher energies as compared to the AFM-G single domain after ionic relaxation; the
FM$_X$ configuration shown in Fig.\ref{fig4}a), that corresponds to a truly magnetic domain wall (MW) between two AFM-G domains with opposite
polarities, results in the
second lowest DW energy, namely $E_{DW+MW}=\unit[46]{mJ/m^2}$, with an energy increase of $\sim \unit[5]{mJ/m^2}$
with respect to the AFM-G DW. The structural-induced modifications of magnetic
exchanges can then be inferred by mapping total energies --- as obtained by enforcing the AFM-G optimal lattice structure ---
onto a Heisenberg model with normalized spins, $H=\sum_{ij}J_{ij}S_iS_j$.
We assumed nearest-neighbor ($nn$) interactions $J_x$, $J_y$, $J_z$ and isotropic next-nearest-neighbor $nnn$
interaction $J_2$ across each interface in the supercell,  plus a $nn$ exchange $J'_x$ between two Mn ions
belonging to first and second MnO$_2$ layers from the twin boundary.
\begin{table}[ht]
\centering
\caption{Heisenberg exchange
coupling constants (in $\unit{meV}$) and corresponding structural informations, $i.e.$ Mn-O-Mn bond angle $\theta$ and Mn-O bond length $d$. Bulk values are also reported in brackets, in qualitative good agreement with previous findings\cite{Nicastro_Patterson}.}
\label{tab1}
\begin{tabular}{p{1.cm}p{1.3cm}p{1.3cm}p{1.3cm}p{1.3cm}p{1.2cm}}
\hline
\centering & \centering{$J_x$}&\centering  $J_y$& \centering $J_z$& \centering $J'_x$& $J_{2}$\\
\hline
& \centering{11.13} & \centering 18.34 &\centering  18.42 &\centering  7.61 & 1.63\\[-0.1cm]
&\centering{\footnotesize (7.44) }&\centering{\footnotesize (7.44) }&\centering{\footnotesize (10.79) }&\centering{\footnotesize (7.44) }&{\footnotesize (1.80) }\\
\hline
\centering $\theta ~(\degree)$&\centering 158.2&\centering 160.5&\centering 158.7&\centering 157.7& \\[-0.1cm]
&\centering {\footnotesize (156.3) }&\centering {\footnotesize (156.3) }&\centering {\footnotesize (158.1) }&\centering {\footnotesize (156.3) }&\\
\centering $d ~$(\AA) & \centering 1.91 & \centering 1.9-1.91 & \centering 1.89-1.91& \centering 1.92 &\\[-0.1cm]
&\centering {\footnotesize (1.92) }&\centering {\footnotesize (1.92) }&\centering {\footnotesize (1.91) }&\centering {\footnotesize (1.92) }&\\
\hline
\end{tabular}
\end{table} 
Our results (see Table \ref{tab1}) show a trend
which agrees with the one previously reported for $Pnma$ perovskites
BiFeO$_3$ and LaFeO$_3$ \cite{Weingart_Spaldin_Bousquet}. As 
the Mn-O-Mn angle $\theta$ is reduced when moving away from the DW,
the exchange coupling component perpendicular to the twin wall
decreases ($J'_x<J_x$). More generally, the structural-induced strong modifications observed in the anisotropic
exchange constants can be qualitatively understood assuming $J\propto t_{pd}^4\cos^2\theta/[\Delta^2(2\Delta+U_{oxy})]$,
where $t_{pd}$
describes the overlap integral between Mn-$d$ and O$-p$ states, $\Delta$ is the $d^3\rightarrow d^4\underline L$ charge
transfer energy and $U_{oxy}$ the on-site correlation energy on O ions\cite{Millis,Meskine}. 
Exchange couplings are strongly affected by both the Mn-O-Mn angle (increasing as $\theta\rightarrow180\degree$) and the Mn-O bond length $d$,
being
$t_{pd}\propto d^{-g}$ and $g=3.5$\cite{Harrison}. From this parametrization, an
offcentering of Mn ions is also expected to enhance the corresponding exchange interaction; in fact,
assuming that a distortion $u$ induces a hybridization change $\Delta t_{pd}\sim\pm g u+ g(g+1)u^2/2$,
one immediately finds that $J\propto t_-^2t_+^2\sim t_{pd}^4(1+g u^2)$.

On the other hand, different local spin configurations at the boundary strongly influence the structural deformations, hence
the polarization profiles. Indeed, in the presence of both a magnetic wall (the so-called
FM$_X$ configuration)and a structural twin boundary, the
offcentering of relaxed Mn ions is strongly reduced along the $Y$ direction, while it is almost doubled along the $X$ direction, in
order to decrease the magnetic energy cost of having parallel spin bonds across the domain wall [see Fig. \ref{fig4}(b)].
Interestingly, this kind of spin-phonon coupling is expected to show up also in the presence of a MW
in a structural (ferroelastic) monodomain; in fact, we predict an antiferroelectric-like profile of Mn off-centerings at
the boundary enforcing the FM$_X$ spin configuration on top of a bulk lattice structure, with no twin boundaries
[see Fig. \ref{fig4}(c)]. Furthermore, the corresponding MW energy $E_{MW}\sim\unit[7.2]{mJ/m^2}$ is slightly larger than the additional energy cost of having the magnetic wall pinned at the twin wall. We additionally note that,
unfortunately, almost all the considered spin configurations
do not display a net interface magnetization, due to perfect compensation of magnetic moments along
different directions. On the other hand, a truly
ferromagnetic (FM) interface, that in principle could be moved and controlled by applying an external magnetic
field, is realized in the FM spin configuration, with all FM bonds around the
boundary, with an estimated rise of the energy of $\sim 0.7$ eV per layer. However, an alternative possibility to engineer a
FM ferroelectric DW, that is left for future analysis, would be that of having a
local canting of spins at the wall, giving rise to weak FM moments.

\section{Conclusions}
In conclusion, we have performed an accurate analysis from first principles of (multi)ferroic domain walls in prototypical orthorhombic
perovskites CaTiO$_3$
and CaMnO$_3$. We have shown that twin boundaries and
domain walls, which naturally occur in ferroic materials, can indeed host features which do not appear in the corresponding
bulk. Specifically, we propose that the predicted and observed twin-wall polarization in CTO has a hybrid improper
origin arising from a cooperative interplay of rotational distortions acting on both the A-site and B-site cations. In this picture, the existence of a secondary
FE instability is substantially reflected only in the anomalous Born effective charges of B-site ions, which cause a significant contribution to wall ${\bm P}$ even for small offcenterings.
However, we argue that DW polar ionic displacements
are primarily determined by two interfacing antipolar structures with different polarities, and as such they are strongly
locked to the local pattern of AFD distortions. In confirmation of this scenario, we found that similar DW features, i.e. a significant offcenter of cations La and Fe along the supercell $Y$ axis,
develop in LaFeO$_3$, which displays in the bulk the same AFD distortions as CTO and CMO but no FE instability involving
Fe-ions offcentering. Furthermore, a very recent TEM experiment unveiled the existence of polar distortions at antiphase boundaries of PbZrO$_3$, an antiferroelectric orthorhombic perovskite with similar rotation/tilting distortions, which have been explained in terms of a general Landau theory approach to ferroelectricity at antiferroelectric domain walls\cite{Wei_nature2014}.
Finally, our results suggest that the coexistence of ferroelectricity and magnetism as arising by the very same ions is indeed possible,
putting forward CMO as a possible material where
multiferroic nanoscopic features can appear at its twin walls, and eventually suggesting a new route to engineer
multiferroicity in single-phase materials. Even though the DW ferroic properties do not seem to be switchable, due to their improper origin, they can in principle provide additional handles to control and move domain walls beside conventional mechanical-based mechanisms.

We thank Dr. J. \'I\~niguez for fruitful discussions and useful insights about cooperative oxygen rotations and
antiferroelectric distortions.
We acknowledge the  MIUR-PRIN project {\em ``Interfacce di ossidi: nuove proprieta' emergenti, multifunzionalita' e
dispositivi per l'elettronica e l'energia (OXIDE)"}, and
PRACE for awarding us access to resource MareNostrum based
in Spain at Barcelona Supercomputing Center (BSC-CNS).


\begin{thebibliography}{35}
\expandafter\ifx\csname natexlab\endcsname\relax\def\natexlab#1{#1}\fi
\expandafter\ifx\csname bibnamefont\endcsname\relax
  \def\bibnamefont#1{#1}\fi
\expandafter\ifx\csname bibfnamefont\endcsname\relax
  \def\bibfnamefont#1{#1}\fi
\expandafter\ifx\csname citenamefont\endcsname\relax
  \def\citenamefont#1{#1}\fi
\expandafter\ifx\csname url\endcsname\relax
  \def\url#1{\texttt{#1}}\fi
\expandafter\ifx\csname urlprefix\endcsname\relax\def\urlprefix{URL }\fi
\providecommand{\bibinfo}[2]{#2}
\providecommand{\eprint}[2][]{\url{#2}}

\bibitem[{\citenamefont{Wada et~al.}(2006)\citenamefont{Wada, Yako, Yokoo,
  Kakemoto, and Tsurumi}}]{Wada2006}
\bibinfo{author}{\bibfnamefont{S.}~\bibnamefont{Wada}},
  \bibinfo{author}{\bibfnamefont{K.}~\bibnamefont{Yako}},
  \bibinfo{author}{\bibfnamefont{K.}~\bibnamefont{Yokoo}},
  \bibinfo{author}{\bibfnamefont{H.}~\bibnamefont{Kakemoto}}, \bibnamefont{and}
  \bibinfo{author}{\bibfnamefont{T.}~\bibnamefont{Tsurumi}},
  \bibinfo{journal}{Ferroelectrics} \textbf{\bibinfo{volume}{{\bf334}}},
  \bibinfo{pages}{17} (\bibinfo{year}{2006}).

\bibitem[{\citenamefont{Salje}(2010)}]{Salje2010}
\bibinfo{author}{\bibfnamefont{E.~K.~H.} \bibnamefont{Salje}},
  \bibinfo{journal}{Chem. Phys. Chem.} \textbf{\bibinfo{volume}{{\bf11}}},
  \bibinfo{pages}{940} (\bibinfo{year}{2010}).

\bibitem[{\citenamefont{Salje}(2012)}]{Salje_annrev}
\bibinfo{author}{\bibfnamefont{E.~K.~H.} \bibnamefont{Salje}},
  \bibinfo{journal}{Annu. Rev. Mater. Res.} \textbf{\bibinfo{volume}{{\bf42}}},
  \bibinfo{pages}{265} (\bibinfo{year}{2012}).

\bibitem[{\citenamefont{Catalan et~al.}(2012)\citenamefont{Catalan, Seidel,
  Ramesh, and Scott}}]{Catalan2012}
\bibinfo{author}{\bibfnamefont{G.}~\bibnamefont{Catalan}},
  \bibinfo{author}{\bibfnamefont{J.}~\bibnamefont{Seidel}},
  \bibinfo{author}{\bibfnamefont{R.}~\bibnamefont{Ramesh}}, \bibnamefont{and}
  \bibinfo{author}{\bibfnamefont{J.~F.} \bibnamefont{Scott}},
  \bibinfo{journal}{Rev. Mod. Phys.} \textbf{\bibinfo{volume}{84}},
  \bibinfo{pages}{119} (\bibinfo{year}{2012}).

\bibitem[{\citenamefont{Lee et~al.}(2003{\natexlab{a}})\citenamefont{Lee,
  Salje, and Bismayer}}]{Lee_Salje_Bismayer1}
\bibinfo{author}{\bibfnamefont{W.~T.} \bibnamefont{Lee}},
  \bibinfo{author}{\bibfnamefont{E.~K.~H.} \bibnamefont{Salje}},
  \bibnamefont{and} \bibinfo{author}{\bibfnamefont{U.}~\bibnamefont{Bismayer}},
  \bibinfo{journal}{J. Appl. Phys.} \textbf{\bibinfo{volume}{{\bf93}}},
  \bibinfo{pages}{9890} (\bibinfo{year}{2003}{\natexlab{a}}).

\bibitem[{\citenamefont{Lee et~al.}(2003{\natexlab{b}})\citenamefont{Lee,
  Salje, and Bismayer}}]{Lee_Salje_Bismayer2}
\bibinfo{author}{\bibfnamefont{W.~T.} \bibnamefont{Lee}},
  \bibinfo{author}{\bibfnamefont{E.~K.~H.} \bibnamefont{Salje}},
  \bibnamefont{and} \bibinfo{author}{\bibfnamefont{U.}~\bibnamefont{Bismayer}},
  \bibinfo{journal}{J. Phys. Condens. Matter}
  \textbf{\bibinfo{volume}{{\bf15}}}, \bibinfo{pages}{1353}
  (\bibinfo{year}{2003}{\natexlab{b}}).

\bibitem[{\citenamefont{Salje}(1993)}]{Salje}
\bibinfo{author}{\bibfnamefont{E.~K.~H.} \bibnamefont{Salje}},
  \emph{\bibinfo{title}{Phase Transition in Ferroelastic and Co-Elastic
  Crystals}} (\bibinfo{publisher}{Cambridge University Press, Cambridge, UK},
  \bibinfo{year}{1993}).

\bibitem[{\citenamefont{Goncalves-Ferreira
  et~al.}(2008)\citenamefont{Goncalves-Ferreira, Redfern, Artacho, and
  Salje}}]{Goncalves-Ferreira_Redfern_Artacho_Salje}
\bibinfo{author}{\bibfnamefont{L.}~\bibnamefont{Goncalves-Ferreira}},
  \bibinfo{author}{\bibfnamefont{S.~A.~T.} \bibnamefont{Redfern}},
  \bibinfo{author}{\bibfnamefont{E.}~\bibnamefont{Artacho}}, \bibnamefont{and}
  \bibinfo{author}{\bibfnamefont{E.~K.~H.} \bibnamefont{Salje}},
  \bibinfo{journal}{Phys. Rev. Lett.} \textbf{\bibinfo{volume}{{\bf101}}},
  \bibinfo{pages}{097602} (\bibinfo{year}{2008}).

\bibitem[{\citenamefont{Cockayne and Burton}(2000)}]{Cockayne_Burton}
\bibinfo{author}{\bibfnamefont{E.}~\bibnamefont{Cockayne}} \bibnamefont{and}
  \bibinfo{author}{\bibfnamefont{B.~P.} \bibnamefont{Burton}},
  \bibinfo{journal}{Phys. Rev. B} \textbf{\bibinfo{volume}{{\bf62}}},
  \bibinfo{pages}{3735} (\bibinfo{year}{2000}).

\bibitem[{\citenamefont{Bhattacharjee et~al.}(2009)\citenamefont{Bhattacharjee,
  Bousquet, and Ghosez}}]{Ghosez_Bousquet}
\bibinfo{author}{\bibfnamefont{S.}~\bibnamefont{Bhattacharjee}},
  \bibinfo{author}{\bibfnamefont{E.}~\bibnamefont{Bousquet}}, \bibnamefont{and}
  \bibinfo{author}{\bibfnamefont{P.}~\bibnamefont{Ghosez}},
  \bibinfo{journal}{Phys. Rev. Lett.} \textbf{\bibinfo{volume}{102}},
  \bibinfo{pages}{117602} (\bibinfo{year}{2009}).

\bibitem[{\citenamefont{Barone et~al.}(2011)\citenamefont{Barone, Kanungo,
  Picozzi, and Saha-Dasgupta}}]{Barone_CMO}
\bibinfo{author}{\bibfnamefont{P.}~\bibnamefont{Barone}},
  \bibinfo{author}{\bibfnamefont{S.}~\bibnamefont{Kanungo}},
  \bibinfo{author}{\bibfnamefont{S.}~\bibnamefont{Picozzi}}, \bibnamefont{and}
  \bibinfo{author}{\bibfnamefont{T.}~\bibnamefont{Saha-Dasgupta}},
  \bibinfo{journal}{Phys. Rev. B} \textbf{\bibinfo{volume}{84}},
  \bibinfo{pages}{134101} (\bibinfo{year}{2011}).

\bibitem[{\citenamefont{Rondinelli et~al.}(2009)\citenamefont{Rondinelli,
  Eidelson, and Spaldin}}]{rondinelli_prb}
\bibinfo{author}{\bibfnamefont{J.~M.} \bibnamefont{Rondinelli}},
  \bibinfo{author}{\bibfnamefont{A.~S.} \bibnamefont{Eidelson}},
  \bibnamefont{and} \bibinfo{author}{\bibfnamefont{N.~A.}
  \bibnamefont{Spaldin}}, \bibinfo{journal}{Phys. Rev. B}
  \textbf{\bibinfo{volume}{79}}, \bibinfo{pages}{205119}
  (\bibinfo{year}{2009}).

\bibitem[{\citenamefont{Bersuker}(2012)}]{Bersuker}
\bibinfo{author}{\bibfnamefont{I.~B.} \bibnamefont{Bersuker}},
  \bibinfo{journal}{Phys. Rev. Lett.} \textbf{\bibinfo{volume}{108}},
  \bibinfo{pages}{137202} (\bibinfo{year}{2012}).

\bibitem[{\citenamefont{Hill}(2000)}]{spaldin_hill}
\bibinfo{author}{\bibfnamefont{N.~A.} \bibnamefont{Hill}}, \bibinfo{journal}{J.
  Phys. Chem. B} \textbf{\bibinfo{volume}{104}}, \bibinfo{pages}{6694}
  (\bibinfo{year}{2000}).


\bibitem[{\citenamefont{Sakai et~al.}(2011)\citenamefont{Sakai, Fujioka,
  Fukuda, Okuyama, Hashizume, Kagawa, Nakao, Murakami, Arima, Baron
  et~al.}}]{sakai}
\bibinfo{author}{\bibfnamefont{H.}~\bibnamefont{Sakai}},
  \bibinfo{author}{\bibfnamefont{J.}~\bibnamefont{Fujioka}},
  \bibinfo{author}{\bibfnamefont{T.}~\bibnamefont{Fukuda}},
  \bibinfo{author}{\bibfnamefont{D.}~\bibnamefont{Okuyama}},
  \bibinfo{author}{\bibfnamefont{D.}~\bibnamefont{Hashizume}},
  \bibinfo{author}{\bibfnamefont{F.}~\bibnamefont{Kagawa}},
  \bibinfo{author}{\bibfnamefont{H.}~\bibnamefont{Nakao}},
  \bibinfo{author}{\bibfnamefont{Y.}~\bibnamefont{Murakami}},
  \bibinfo{author}{\bibfnamefont{T.}~\bibnamefont{Arima}},
  \bibinfo{author}{\bibfnamefont{A.~Q.~R.} \bibnamefont{Baron}},
  \bibnamefont{et~al.}, \bibinfo{journal}{Phys. Rev. Lett.}
  \textbf{\bibinfo{volume}{107}}, \bibinfo{pages}{137601}
  (\bibinfo{year}{2011}).

\bibitem[{\citenamefont{G\"unter et~al.}(2012)\citenamefont{G\"unter, Bousquet,
  David, Boullay, Gho~sez, Prellier, and Fiebig}}]{cmo_film}
\bibinfo{author}{\bibfnamefont{T.}~\bibnamefont{G\"unter}},
  \bibinfo{author}{\bibfnamefont{E.}~\bibnamefont{Bousquet}},
  \bibinfo{author}{\bibfnamefont{A.}~\bibnamefont{David}},
  \bibinfo{author}{\bibfnamefont{P.}~\bibnamefont{Boullay}},
  \bibinfo{author}{\bibfnamefont{P.}~\bibnamefont{Gho~sez}},
  \bibinfo{author}{\bibfnamefont{W.}~\bibnamefont{Prellier}}, \bibnamefont{and}
  \bibinfo{author}{\bibfnamefont{M.}~\bibnamefont{Fiebig}},
  \bibinfo{journal}{Phys. Rev. B} \textbf{\bibinfo{volume}{85}},
  \bibinfo{pages}{214120} (\bibinfo{year}{2012}).

\bibitem[{\citenamefont{{Van Aert} et~al.}(2012)\citenamefont{{Van Aert},
  Turner, Delville, Schryvers, {Van Tendeloo}, and
  Salje}}]{VanAert_Turner_Delville_Schryvers_VanTendeloo_Salje}
\bibinfo{author}{\bibfnamefont{S.}~\bibnamefont{{Van Aert}}},
  \bibinfo{author}{\bibfnamefont{S.}~\bibnamefont{Turner}},
  \bibinfo{author}{\bibfnamefont{R.}~\bibnamefont{Delville}},
  \bibinfo{author}{\bibfnamefont{D.}~\bibnamefont{Schryvers}},
  \bibinfo{author}{\bibfnamefont{G.}~\bibnamefont{{Van Tendeloo}}},
  \bibnamefont{and} \bibinfo{author}{\bibfnamefont{E.~K.~H.}
  \bibnamefont{Salje}}, \bibinfo{journal}{Adv. Mater.}
  \textbf{\bibinfo{volume}{{\bf24}}}, \bibinfo{pages}{523}
  (\bibinfo{year}{2012}).

\bibitem[{\citenamefont{Benedek and Fennie}(2011)}]{benedek_prl}
\bibinfo{author}{\bibfnamefont{N.~A.} \bibnamefont{Benedek}} \bibnamefont{and}
  \bibinfo{author}{\bibfnamefont{C.~J.} \bibnamefont{Fennie}},
  \bibinfo{journal}{Phys. Rev. Lett.} \textbf{\bibinfo{volume}{106}},
  \bibinfo{pages}{107204} (\bibinfo{year}{2011}).

\bibitem[{\citenamefont{Rondinelli and Fennie}(2012)}]{rondinelli_advmat}
\bibinfo{author}{\bibfnamefont{J.~M.} \bibnamefont{Rondinelli}}
  \bibnamefont{and} \bibinfo{author}{\bibfnamefont{C.~J.}
  \bibnamefont{Fennie}}, \bibinfo{journal}{Adv. Mater.}
  \textbf{\bibinfo{volume}{24}}, \bibinfo{pages}{1961} (\bibinfo{year}{2012}).


\bibitem[{\citenamefont{Bellaiche and \'I\~niguez}(2013)}]{Jorge}
\bibinfo{author}{\bibfnamefont{L.}~\bibnamefont{Bellaiche}} \bibnamefont{and}
  \bibinfo{author}{\bibfnamefont{J.}~\bibnamefont{\'I\~niguez}},
  \bibinfo{journal}{Phys. Rev. B} \textbf{\bibinfo{volume}{88}},
  \bibinfo{pages}{014104} (\bibinfo{year}{2013}).








\bibitem[{\citenamefont{Perdew et~al.}(2008)\citenamefont{Perdew, Ruzsinszky,
  Csonka, Vydrov, Scuseria, Constantin, Zhou, and
  Burke}}]{Perdew_Ruzsinszky_Csonka_Vydrov_Scuseria_Constantin_Zhou_Burke}
\bibinfo{author}{\bibfnamefont{J.~P.} \bibnamefont{Perdew}},
  \bibinfo{author}{\bibfnamefont{A.}~\bibnamefont{Ruzsinszky}},
  \bibinfo{author}{\bibfnamefont{G.~I.} \bibnamefont{Csonka}},
  \bibinfo{author}{\bibfnamefont{O.~A.} \bibnamefont{Vydrov}},
  \bibinfo{author}{\bibfnamefont{G.~E.} \bibnamefont{Scuseria}},
  \bibinfo{author}{\bibfnamefont{L.~A.} \bibnamefont{Constantin}},
  \bibinfo{author}{\bibfnamefont{X.}~\bibnamefont{Zhou}}, \bibnamefont{and}
  \bibinfo{author}{\bibfnamefont{K.}~\bibnamefont{Burke}},
  \bibinfo{journal}{Phys. Rev. Lett.} \textbf{\bibinfo{volume}{{\bf100}}},
  \bibinfo{pages}{136406} (\bibinfo{year}{2008}).

\bibitem[{\citenamefont{Kresse and Joubert}(1999)}]{vasp}
\bibinfo{author}{\bibfnamefont{G.}~\bibnamefont{Kresse}} \bibnamefont{and}
  \bibinfo{author}{\bibfnamefont{D.}~\bibnamefont{Joubert}},
  \bibinfo{journal}{Phys. Rev. B} \textbf{\bibinfo{volume}{{\bf59}}},
  \bibinfo{pages}{1758} (\bibinfo{year}{1999}).

\bibitem[{\citenamefont{Matar}(2003)}]{Matar}
\bibinfo{author}{\bibfnamefont{S.~F.} \bibnamefont{Matar}},
  \bibinfo{journal}{Prog. Solid State Chem.} \textbf{\bibinfo{volume}{31}},
  \bibinfo{pages}{239} (\bibinfo{year}{2003}).

\bibitem[{\citenamefont{Glazer}(1972)}]{Glazer}
\bibinfo{author}{\bibfnamefont{A.~M.} \bibnamefont{Glazer}},
  \bibinfo{journal}{Acta Cryst.} \textbf{\bibinfo{volume}{{\bf B28}}},
  \bibinfo{pages}{3384} (\bibinfo{year}{1972}).

\bibitem[{\citenamefont{Gopalan}(1972)}]{Gopalan}
\bibinfo{author}{\bibfnamefont{V.} \bibnamefont{Gopalan}} \bibnamefont{and}
\bibinfo{author}{\bibfnamefont{D.~B.} \bibnamefont{Litvin}},
  \bibinfo{journal}{Nature Mater.} \textbf{\bibinfo{volume}{{\bf 10}}},
  \bibinfo{pages}{376} (\bibinfo{year}{2011}).



\bibitem{note2}The DW
energies are computed as $E_{DW}=(E-E_0)/2S$, where $E$ is the energy of the DW configuration, $E_0$
the energy of the bulk structure evaluated for the same supercell and $S$ is the surface area of the cell parallel
to the DW. The factor 2 comes from the presence of two DWs in the supercell.

\bibitem{footnote_extra} In other words, rotations around the pseudocubic $X$ axis appears to be in-phase just at the DW, whereas rotations around pseudocubic axes $Y, Z$ remain out-of-phase and in-phase, respectively; the local pattern of rotations around the DW would correspond then to a $b^+a^-c^+$ Glazer's configuration. 

\bibitem{vanderbilt_PTO} B. Meyer and D. Vanderbilt, Phys. Rev. B {\bf 65}, 104111 (2002).


\bibitem{note3} In order to further verify the B ions offcentering caused by the AFD distortions, we also performed a calculation where the A ions have been fixed to their centrosymmetric positions (without antipolar distortions) and oxygens to their optimal positions (i.e. in the ground-state rotational pattern): when B ions are allowd to relax from centrosymmetric position, they adopt a distorted pattern qualitatively consistent with the fully optimized structure, even though their absolute offcentering is reduced to $\sim \unit[1]{pm}$ at the CTO DW.


\bibitem[{\citenamefont{Hong et~al.}(2012)\citenamefont{Hong, Stroppa,
  \'I\~niguez, Picozzi, and Vanderbilt}}]{Hong_Stroppa_Iniguez}
\bibinfo{author}{\bibfnamefont{J.}~\bibnamefont{Hong}},
  \bibinfo{author}{\bibfnamefont{A.}~\bibnamefont{Stroppa}},
  \bibinfo{author}{\bibfnamefont{J.}~\bibnamefont{\'I\~niguez}},
  \bibinfo{author}{\bibfnamefont{S.}~\bibnamefont{Picozzi}}, \bibnamefont{and}
  \bibinfo{author}{\bibfnamefont{D.}~\bibnamefont{Vanderbilt}},
  \bibinfo{journal}{Phys. Rev. B} \textbf{\bibinfo{volume}{85}},
  \bibinfo{pages}{054417} (\bibinfo{year}{2012}).


\bibitem[{\citenamefont{Nicastro and Patterson}(2002)}]{Nicastro_Patterson}
\bibinfo{author}{\bibfnamefont{M.}~\bibnamefont{Nicastro}} \bibnamefont{and}
  \bibinfo{author}{\bibfnamefont{C.~H.} \bibnamefont{Patterson}},
  \bibinfo{journal}{Phys. Rev. B} \textbf{\bibinfo{volume}{65}},
  \bibinfo{pages}{205111} (\bibinfo{year}{2002}).

\bibitem[{\citenamefont{Weingart et~al.}(2012)\citenamefont{Weingart, Spaldin,
  and Bousquet}}]{Weingart_Spaldin_Bousquet}
\bibinfo{author}{\bibfnamefont{C.}~\bibnamefont{Weingart}},
  \bibinfo{author}{\bibfnamefont{N.}~\bibnamefont{Spaldin}}, \bibnamefont{and}
  \bibinfo{author}{\bibfnamefont{E.}~\bibnamefont{Bousquet}},
  \bibinfo{journal}{arXiv:1206.0718v1 [cond-mat.mtrl-sci]}
  (\bibinfo{year}{2012}).

\bibitem[{\citenamefont{Millis}(1997)}]{Millis}
\bibinfo{author}{\bibfnamefont{A.~J.} \bibnamefont{Millis}},
  \bibinfo{journal}{Phys. Rev. B} \textbf{\bibinfo{volume}{55}},
  \bibinfo{pages}{6405} (\bibinfo{year}{1997}).

\bibitem[{\citenamefont{Meskine et~al.}(2001)\citenamefont{Meskine, K\"onig,
  and Satpathy}}]{Meskine}
\bibinfo{author}{\bibfnamefont{H.}~\bibnamefont{Meskine}},
  \bibinfo{author}{\bibfnamefont{H.}~\bibnamefont{K\"onig}}, \bibnamefont{and}
  \bibinfo{author}{\bibfnamefont{S.}~\bibnamefont{Satpathy}},
  \bibinfo{journal}{Phys. Rev. B} \textbf{\bibinfo{volume}{64}},
  \bibinfo{pages}{094433} (\bibinfo{year}{2001}).

\bibitem[{\citenamefont{Harrison}(1980)}]{Harrison}
\bibinfo{author}{\bibfnamefont{W.~A.} \bibnamefont{Harrison}},
  \emph{\bibinfo{title}{Electronic Structure and the Properties of Solids}}
  (\bibinfo{publisher}{Freeman, San Francisco, USA}, \bibinfo{year}{1980}).

\bibitem[{\citenamefont{Wei}(2014)}]{Wei_nature2014}
\bibinfo{author}{\bibfnamefont{X.-K.} \bibnamefont{Wei}},
 \bibinfo{author}{\bibfnamefont{A.~K.} \bibnamefont{Tagantsev}},
 \bibinfo{author}{\bibfnamefont{A.} \bibnamefont{Kvasov}}, 
 \bibinfo{author}{\bibfnamefont{K.} \bibnamefont{Roleder}}, 
 \bibinfo{author}{\bibfnamefont{C.-L.} \bibnamefont{Jia}} \bibnamefont{and}
  \bibinfo{author}{\bibfnamefont{N.}~\bibnamefont{Setter}},
  \bibinfo{journal}{Nat. Commun.} \textbf{\bibinfo{volume}{5}},
  \bibinfo{pages}{3031} (\bibinfo{year}{2014}).
  

\end{thebibliography}

\end{document}